\RequirePackage{fix-cm} 
\documentclass[a4paper, twoside, reqno, 12pt]{amsart}
\usepackage{fixltx2e}   

\usepackage[latin1]{inputenc}
\usepackage[T1]{fontenc}

\usepackage{eucal}
\usepackage{esint}
\usepackage{dsfont}
\usepackage{xspace}
\usepackage{amsgen}
\usepackage{amsthm}
\usepackage{amssymb}
\usepackage{amsmath}
\usepackage{upgreek}
\usepackage{amsfonts}
\usepackage{MnSymbol}
\usepackage{mathrsfs}
\usepackage{mathtools}
\usepackage[nice]{nicefrac}

\usepackage{a4wide}

\usepackage[table]{xcolor}

\definecolor{oneblue}{rgb}{0.0, 0.0, 0.85}
\definecolor{darkgrey}{rgb}{0.273, 0.281, 0.30}
\definecolor{Lightgray}{rgb}{0.89, 0.89, 0.89}
\definecolor{Lightblue}{RGB}{214, 214, 214}
\definecolor{bckg}{RGB}{20.8, 20.8, 20.8} 

\usepackage{psfrag}
\usepackage{graphicx}
\usepackage{subfigure}
\usepackage{morefloats}
\usepackage{indentfirst}

\usepackage{acronym}
\usepackage{microtype}
\usepackage[labelfont={bf,sf},%
            labelsep=period,%
            justification=raggedright]{caption}

\usepackage[usenames, dvipsnames, pdf]{pstricks}
\usepackage{epsfig}
\usepackage{pst-grad} 
\usepackage{pst-plot} 

\usepackage[colorlinks,
            urlcolor=oneblue,
            linkcolor=oneblue,
            citecolor=oneblue,
            bookmarksopen=false,
            pdfpagemode=UseNone,
            pagebackref]{hyperref}

\usepackage[explicit]{titlesec}

\titleformat{\section}
  {\color{darkgrey}\Large\sffamily\bfseries}
  {}
  {0em}
  {\colorbox{bckg!5}{\parbox{\dimexpr\linewidth-2\fboxsep\relax}{\centering\thesection. #1}}}
  []

\titleformat{name=\section,numberless}
  {\color{darkgrey}\Large\sffamily\bfseries}
  {}
  {0em}
  {\colorbox{bckg!10}{\parbox{\dimexpr\linewidth-2\fboxsep\relax}{\centering#1}}}
  []

\titleformat{\subsection}
  {\color{darkgrey}\large\sffamily\bfseries}
  {}
  {0.0em}
  {\colorbox{bckg!5}{\parbox{\dimexpr\linewidth-2\fboxsep\relax}{\centering\thesubsection. #1}}}
  []

\titleformat{name=\subsection,numberless}
  {\color{darkgrey}\Large\sffamily\bfseries}
  {}
  {0em}
  {\colorbox{bckg!10}{\parbox{\dimexpr\linewidth-2\fboxsep\relax}{\centering#1}}}
  []

\titleformat{\subsubsection}
  {\sffamily\normalsize\bfseries}
  {\thesubsubsection}
  {0.5em}
  {#1}
  []

\titleformat{\paragraph}[runin]
  {\sffamily\small\bfseries}
  {}
  {0em}
  {#1}

\titlespacing*{\section}{1.0em}{1.0em}{0.8em}[0em]
\titlespacing*{\subsection}{1.0em}{1.0em}{0.8em}[0em]
\titlespacing*{\subsubsection}{1.0em}{0.7em}{0.6em}[0em]

\usepackage{titletoc}

\contentsmargin{0.5em}
\newlength{\tocsep} 
\titlecontents{section}[\tocsep]
  {\addvspace{10pt}\bfseries\sffamily}
  {\contentslabel[\thecontentslabel]{\tocsep}}
  {}
  {\ \titlerule*[0.75pc]{.}\ \thecontentspage}
  []
\titlecontents{subsection}[\tocsep]
  {\addvspace{8pt}\sffamily}
  {\contentslabel[\thecontentslabel]{\tocsep}}
  {}
  {\ \titlerule*[0.5pc]{.}\ \thecontentspage}
  []
\titlecontents*{subsubsection}[\tocsep]
  {\footnotesize\sffamily}
  {}
  {}
  {\ \titlerule*[0.35pc]{.}\ \thecontentspage}
  [\ \textbullet\ ]  

\usepackage{fancyhdr}
\usepackage{lastpage}

\newcommand*\Title{A new run-up algorithm}
\newcommand*\Authors{G.~Khakimzyanov \etal}
\newcommand*{\plogo}{{\texttt{arXiv.org} / \textsc{hal}}} 

\numberwithin{equation}{section}

\newtheorem{remark}{Remark}
\newtheorem{theorem}{Theorem}
\newtheorem{corollary}{Corollary}

\newcommand{\up}[1]{$^{\mathrm{\small\textsf{#1}}}$} 

\newcommand{\g}{\mathbf{g}}

\newcommand{\A}{\mathscr{A}}
\newcommand{\D}{\mathcal{D}}

\newcommand{\ud}{\mathrm{d}}

\newcommand{\U}{\mathcal{U}}
\newcommand{\Ru}{\mathcal{R}}
\newcommand{\x}{\overline{x}}

\renewcommand{\O}{\mathcal{O}}

\renewcommand{\H}{\mathcal{H}}

\newcommand{\f}{\boldsymbol{f}}
\newcommand{\Ss}{\boldsymbol{S}}
\renewcommand{\v}{\boldsymbol{v}}
\newcommand{\const}{\mathrm{const}}
\newcommand{\C}{\boldsymbol{\mathrm{C}}}
\renewcommand{\gamma}{\boldsymbol{\upgamma}}

\newcommand{\ie}{\emph{i.e.}~}
\newcommand{\eg}{\emph{e.g.}~}
\newcommand{\etal}{\emph{et al.}~}

\newcommand{\sech}{\mathrm{sech}}
\newcommand{\abs}[1]{\left|#1\right|}

\newcommand{\pd}[2]{\frac{\partial #1}{\partial\/ #2}}
\newcommand{\od}[2]{\frac{\mathrm{d} #1}{\mathrm{d}\/#2}}

\usepackage{acronym}
\acrodef{BVP}{Boundary Value Problem}
\acrodef{NSWE}{Nonlinear Shallow Water Equations}

\begin{document}

\title[\Title]{A new run-up algorithm based on local high-order analytic expansions}

\author[G.~Khakimzyanov]{Gayaz Khakimzyanov$^*$}
\address{Institute of Computational Technologies, Siberian Branch of the Russian Academy of Sciences, Novosibirsk 630090, Russia}
\email{Khak@ict.nsc.ru}
\urladdr{http://www.ict.nsc.ru/ru/structure/Persons/ict-KhakimzyanovGS}
\thanks{$^*$ Corresponding author}

\author[N.Yu.~Shokina]{Nina Yu. Shokina}
\address{Institute of Computational Technologies, Siberian Branch of the Russian Academy of Sciences, Novosibirsk 630090, Russia}
\email{Nina.Shokina@googlemail.com}
\urladdr{http://www.ict.nsc.ru/ru/structure/Persons/ict-ShokinaNY}

\author[D.~Dutykh]{Denys Dutykh}
\address{LAMA, UMR 5127 CNRS, Universit\'e Savoie Mont Blanc, Campus Scientifique, 73376 Le Bourget-du-Lac Cedex, France}
\email{Denys.Dutykh@univ-savoie.fr}
\urladdr{http://www.denys-dutykh.com/}

\author[D.~Mitsotakis]{Dimitrios Mitsotakis}
\address{Victoria University of Wellington, School of Mathematics, Statistics and Operations Research, PO Box 600, Wellington 6140, New Zealand}
\email{dmitsot@gmail.com}
\urladdr{http://dmitsot.googlepages.com/}


\begin{titlepage}
\thispagestyle{empty} 
\noindent
{\Large Gayaz \textsc{Khakimzyanov}}\\
{\it\textcolor{gray}{Institute of Computational Technologies, Novosibirsk, Russia}}\\[0.02\textheight]
{\Large Nina Yu. \textsc{Shokina}}\\
{\it\textcolor{gray}{Institute of Computational Technologies, Novosibirsk, Russia}}\\[0.02\textheight]
{\Large Denys \textsc{Dutykh}}\\
{\it\textcolor{gray}{CNRS, Universit\'e Savoie Mont Blanc, France}}
\\[0.02\textheight]
{\Large Dimitrios \textsc{Mitsotakis}}\\
{\it\textcolor{gray}{Victoria University of Wellington, New Zealand}}\\[0.16\textheight]

\colorbox{Lightblue}{
  \parbox[t]{1.0\textwidth}{
    \centering\huge\sc
    \vspace*{0.7cm}
    
    A new run-up algorithm based on local high-order analytic expansions

    \vspace*{0.7cm}
  }
}

\vfill 

\raggedleft     
{\large \plogo} 
\end{titlepage}


\newpage
\maketitle
\thispagestyle{empty}


\begin{abstract}
The practically important problem of the wave run-up is studied in this article in the framework of \acf{NSWE}. The main novelty consists in the usage of high order local asymptotic analytical solutions in the vicinity of the shoreline. Namely, we use the analytical techniques introduced by S.~\textsc{Kovalevskaya} and the analogy with the compressible gas dynamics (\ie gas outflow problem into the vacuum). Our run-up algorithm covers all the possible cases of the wave slope on the shoreline and it incorporates the new analytical information in order to determine the shoreline motion to higher accuracy. The application of this algorithm is illustrated on several important practical examples. Finally, the simulation results are compared with the well-known analytical and experimental predictions.

\bigskip
\noindent \textbf{\keywordsname:} Nonlinear shallow water equations; finite volumes; finite differences; wave run-up; asymptotic expansion \\

\smallskip
\noindent \textbf{MSC:} \subjclass[2010]{74J15 (primary), 74S10, 74J30 (secondary)}

\end{abstract}

\newpage
\tableofcontents
\thispagestyle{empty}


\newpage
\section{Introduction}

The wave run-up problem has been attracting a lot of attention of the hydrodynamicists, coastal engineers and applied mathematicians because of its obvious practical importance for the assessment of inundation maps and mitigation of natural hazards \cite{Titov1997}. Most oftenly this problem has been approached in the context of \acf{NSWE} which was successfully validated several times \cite{Synolakis1987, Synolakis2008}.

Various linearized theories have been applied to estimate the wave run-up \cite{Didenkulova2008}. Perhaps, the most outstanding method is the widely-known Carrier-Greenspan hodograph transformation which allows to transform the \acs{NSWE} into a linear wave equation \cite{CG58}. Later, this method was extended to solve also the \acf{BVP} for the \acs{NSWE} \cite{Antuono2007}. The general conclusion to which several authors have converged is that the linear theory is able to predict correctly the maximal wave run-up \cite{Synolakis1991}. However, this technique has at least one serious shortcoming since it is limited only to constant sloping beaches. Therefore, the usage of numerical techniques seems to be unavoidable \cite{Medeiros2013}.

The first tentatives to simulate numerically the run-up date back to the mid 70's \cite{Lyatkher1974}. Various techniques have been tested ranging from the analytical mappings to a fixed computational domain to the application of moving grids. The most widely employed technique consists in replacing the dry area by a thin water layer of negligible height (see \cite{Kobayashi1989} among the others). Perhaps, the first \emph{modern} numerical run-up algorithm was proposed by \textsc{Hibberd} \& \textsc{Peregrine} (1979) \cite{Hibberd1979}.

The run-up algorithm proposed in the present study is based on two ingredients. The first trick consists in discretizing the fluid domain solely which allows us to use judiciously the grid points only where they are needed in contrast to shock-capturing schemes where the whole computational domain (wet $\bigcup$ dry areas) is discretized. The other ingredient consists in using a high order asymptotic expansion near the moving shoreline point. To the lowest order these solutions can be identified with the so-called \emph{shoreline Riemann problem} \cite{Stoker1958, Bellotti2001, Brocchini2001}. However, our asymptotic solutions are valid not only for flat, but also for general bottoms. In some particular cases they can provide us with the \emph{exact} solution when it is a polynomial function in time. This novel analytical tool allows us to make a zoom on the \acs{NSWE} solutions structure locally in time in a wider class of physical situations. This algorithm is completely unknown outside the Russian literature \cite{Bautin2011}, which justifies the present publication. Moreover, in the present study we focus on the motion of the shoreline which is used in the run-up algorithm, instead of obtaining the solutions local in time in the whole domain.

The key idea to derive and apprehend these asymptotics lies in the deep analogy between the \acs{NSWE} and the compressible Euler equations for an ideal polytropic gas. Another similarity consists in the analogy between the wave run-up (wetting/drying) process and the compressible gas outflow into the vacuum. In this case, the shoreline can be identified with the vacuum boundary. Consequently, one can hope to transpose the powerful analytical methods of the compressible gas dynamics \cite{Bautin2005, Bautin2009} to \acs{NSWE}. This programme will be accomplished hereinbelow.

This paper is organized as follows. In Section~\ref{sec:model} the governing equations are presented and some of their basic properties are discussed. The following Section~\ref{sec:sol} presents a novel high order asymptotic solution in the vicinity of the shoreline point. The numerical run-up algorithm based on this solution is described in Section~\ref{sec:algo} and some numerical results are presented in Section~\ref{sec:res}. Finally, the main conclusions and perspectives of this study are outlined in Section~\ref{sec:concl}.


\section{Mathematical model}\label{sec:model}

\begin{figure}
  \centering
  \includegraphics[width = 0.75\textwidth]{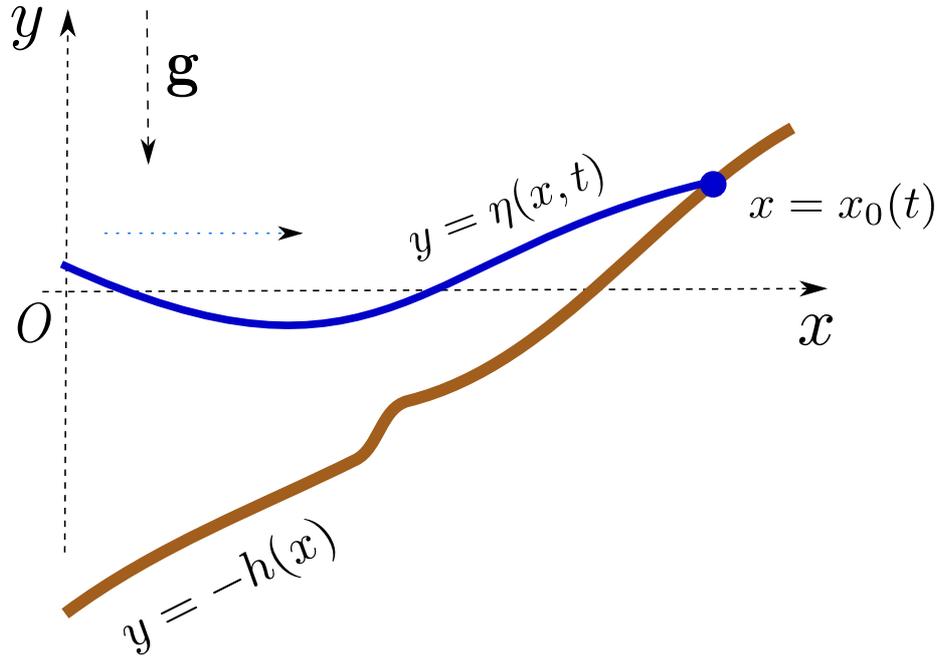}
  \caption{\small\em Sketch of the physical domain, where $y = -h(x)$ is the bottom shape, $y = \eta(x,t)$ is the free surface and $x = x_0(t)$ is the moving shoreline. The horizontal blue dotted arrow shows the wave propagation direction.}
  \label{fig:sketch}
\end{figure}

Consider an ideal and incompressible fluid bounded from below by the absolutely rigid solid bottom (given by the equation $y = -h(x)$) and by the free surface on the top (given by $y = \eta(x,t)$). The sketch of the physical domain and the description of the chosen coordinate system are given on Figure~\ref{fig:sketch}. The Cartesian coordinate system $xOy$ is chosen such that the horizontal axis $y = 0$ coincides with the mean water level (\ie the undisturbed position). The vector $\g = (0, -g)$ denotes the gravity acceleration. The fluid domain is bounded on the right by a sloping beach and $x = x_0(t)$ denotes the instantaneous position of the shoreline.

If we make an additional assumption on the shallowness of the gravity wave propagation (\ie the characteristic wavelength is much bigger than the mean water depth) then it can be shown that the fluid flow is described by the classical \acf{NSWE} \cite{SV1871, Stoker1958}:
\begin{equation}\label{eq:sv}
  \pd{\v}{t}\ +\ \pd{\f(\v)}{x}\ =\ \Ss(\v,x), \quad x\, <\, x_0(t), \quad t\, >\, 0.
\end{equation}
The vector $\v$ of conservative variables, the flux function $\f(\v)$ and the source term are given by the following formulas
\begin{equation*}
  \v\ =\ \begin{pmatrix}
           H \\
           H u
         \end{pmatrix}, \qquad
  \f(\v) =\ \begin{pmatrix}
           H u \\
           H u^2\ +\ \frac{g}{2}H^2
         \end{pmatrix}, \qquad
  \Ss(\v,x)\ =\ \begin{pmatrix}
           0 \\
           gH h_x
         \end{pmatrix}.
\end{equation*}
Physically, the variable $u(x,t)$ is the depth-averaged horizontal velocity and $H(x,t) = h(x) + \eta(x,t)$ is the total water depth.

Equations \eqref{eq:sv} have to be supplied by appropriate initial and boundary conditions. At the moment of time $t = t_0$ we assume that we know the fluid state
\begin{equation}\label{eq:ic}
  H(x,t_0)\ =\ H_0(x), \qquad u(x,t_0)\ =\ u_0(x), \qquad x\, <\, x_0(t_0).
\end{equation}
On the shoreline the total water depth is known
\begin{equation}\label{eq:bc}
  H\bigl(x_0(t), t\bigr)\ =\ 0, \qquad \forall t\, \geqslant\, 0.
\end{equation}
In \cite{Hibberd1979} an additional (to \eqref{eq:bc}) boundary condition on the wet/dry interface has been imposed:
\begin{equation}\label{eq:bc2}
  \od{x_0}{t}\ =\ u\bigl(x_0(t),t\bigr), \qquad \forall t\, \geqslant\, 0,
\end{equation}
however this condition is a direct consequence of \eqref{eq:bc} and it is not required in the present study. Indeed, taking the time derivative of \eqref{eq:bc} and using the mass conservation equation from \eqref{eq:sv} yields straightforwardly \eqref{eq:bc2}. The computational domain will be bounded from the left at $x = 0$ where a non-reflective boundary condition is imposed.

\subsection{Properties}

Equations \eqref{eq:sv} can be recast in the following non-conservative form
\begin{equation*}
  \v_t\ +\ \A\v_x\ =\ \Ss(\v, x),
\end{equation*}
where $\A(\v) := \od{\f(\v)}{\v}$ is the Jacobian matrix. Eigenvalues of the matrix $\A$ can be easily computed
\begin{equation*}
  \lambda_{\pm}\ =\ u\ \pm\ c_s, \qquad c_s\ :=\ \sqrt{gH}.
\end{equation*}
For $x < x_0(t)$ the total depth $H(x,t) > 0$ is necessarily positive and in this case we have $\lambda_+ \neq \lambda_-$. It means that the system \eqref{eq:sv} is hyperbolic in wet areas. On the shoreline both eigenvalues coincide ($\lambda_+ \equiv \lambda_-$) and the line $x = x_0(t)$ is a characteristic of multiplicity two. This observation shows the deep similarity between the shallow water flows and the compressible gas dynamics. In the latter case the separating boundary with the vacuum is also a multiple characteristic. This analogy will allow us to apply the methods developed for compressible gas dynamics to compute the local asymptotic solutions in the neighborhood of the gas/void transition \cite{Bautin2005}.


\section{Local asymptotic solution}\label{sec:sol}

Consider the \acs{NSWE} system \eqref{eq:sv} in the non-conservative form:
\begin{eqnarray}\label{eq:nc1}
  H_t\ +\ uH_x\ +\ Hu_x\ &=&\ 0, \\
  u_t\ +\ uu_x\ +\ gH_x\ &=&\ gh'(x), \label{eq:nc2}
\end{eqnarray}
supplemented with the initial condition \eqref{eq:ic} along with the shoreline boundary condition \eqref{eq:bc}. In the sequel we assume the functions $H_0(x)$, $u_0(x)$ and $h(x)$ to be analytic.

Depending on the initial data $H_0(x)$ there are three distinct possibilities to be analyzed:
\begin{description}
  \item[Regular wave] $H_0'\bigl(x_0(t_0)\bigr) \neq 0$
  \item[Tangent wave] $H_0'\bigl(x_0(t_0)\bigr) = 0$
  \item[Breaking wave\footnotemark] $H_0'\bigl(x_0(t_0)\bigr) = -\infty$
\end{description}
\footnotetext{Since in the shallow water wave theory the wave breaking event is identified with a shock wave formation, it is accompanied by the well-known gradient catastrophe \cite{Kosinski1977} (\ie the classical two-sided derivative tends to infinity).}
The only remaining possibility is $H_0'\bigl(x_0(t_0)\bigr) = +\infty$. However, this situation is not possibe since such a shock wave would not be entropic. In other words, it violates the entropy conditions \cite{Godlewski1990}.

\subsection{Regular wave}\label{sec:case1}

A regular wave contact with a sloping beach is shown for illustration on Figure~\ref{fig:sketch}. By the Theorem of Cauchy--Kovalevskaya \cite{Folland1995} the problem \eqref{eq:nc1}-\eqref{eq:nc2} has a unique analytic solution in the form of the following power series
\begin{equation}\label{eq:kov}
  H(x,t)\ =\ \sum_{k = 0}^{\infty}\;  H_k(x)\; \frac{(t-t_0)^k}{k!}, \qquad u(x,t)\ =\ \sum_{k = 0}^{\infty}\;  u_k(x)\; \frac{(t-t_0)^k}{k!}.
\end{equation}
Now we need to determine the coefficients $H_k(x)$, $u_k(x)$ of this power series. To do so, we substitute the representation \eqref{eq:kov} into the $k$-th derivative in time of the equations \eqref{eq:nc1}-\eqref{eq:nc2} and we evaluate the sum at $t = t_0$. In this way the following recurrence formula can be obtained
\begin{equation*}
  H_{k + 1}(x)\ =\ -\sum_{i = 0}^k\;  \C_k^i\; \bigl(H_i'u_{k-i}\ +\ H_i u_{k-i}'\bigr), \qquad
  u_{k + 1}(x)\ =\ -gH_k'\ -\ \sum_{i = 0}^{k}\; \C_k^i\; u_i' u_{k-i},
\end{equation*}
where $\C_k^i$ are the binomial coefficients. Note that the zeroth order terms $H_0(x)$, $u_0(x)$ are provided directly by the initial conditions \eqref{eq:ic}.

If we assume additionally that the shoreline position is also an analytical function of time, it can be expanded in the Taylor series in powers of $(t-t_0)$
\begin{equation}\label{eq:taylor}
  x_0(t)\ =\ \sum_{k = 0}^{\infty}\;  x^{(k)}_0(t_0)\; \frac{(t-t_0)^k}{k!}, \qquad x^{(k)}_0(t)\ :=\ \od{^k x_0(t)}{t^k}.
\end{equation}
In order to determine the coefficients $x^{(k)}(t_0)$ one needs to substitute \eqref{eq:taylor} into the shoreline boundary condition:
\begin{equation*}
  H\bigl(x_0(t),t\bigr)\ \equiv\ \sum_{k = 0}^{\infty}\;  H_k\bigl(x_0(t)\bigr)\; \frac{(t-t_0)^k}{k!}\ \equiv\ 0.
\end{equation*}
After differentiating the last equality $k$ times with respect to $t$ and evaluating it at $t = t_0$ yields
\begin{equation*}
  x^{(k)}_0(t_0)\ =\ -\frac{H_k\bigl(x_0(t_0)\bigr)\ +\ R_k}{H_0'\bigl(x_0(t_0)\bigr)},
\end{equation*}
where the functions $R_k$ depend recursively on the solution at the preceding orders. First three expressions of $R_k$ are given below
\begin{eqnarray*}
  R_1 &=& 0, \\
  R_2 &=& 2H_1'\bigl(x_0(t_0)\bigr)x^{(1)}_0(t_0)\ +\ H_0''\bigl(x_0(t_0)\bigr)\bigl(x^{(1)}_0(t_0)\bigr)^2, \\
  R_3 &=& 3\bigl(H_1'x^{(2)}_0 + H_2'x^{(1)}_0\bigr)\ +\ 3\bigl(H_0''x^{(2)}_0 + H_1''x^{(1)}_0\bigr)x^{(1)}_0\ +\ H_0''' \bigl(x^{(1)}_0\bigr)^3,
\end{eqnarray*}
where in the last expression for $R_3$ we omitted the argument $x_0(t_0)$ for the sake of conciseness. All coefficients $x^{(k)}_0(t_0)$ can be determined in this way since by assumption we are in a regular situation, \ie $H_0'\bigl(x_0(t_0)\bigr) \neq 0$. Other cases will be treated below.

In order to determine the shoreline velocity $u\bigl(x_0(t), t\bigr)$ let us introduce a new independent variable $\x := x - x_0(t)$ such that a fixed point $\x \equiv 0$ corresponds to the moving shoreline in the initial coordinates. In new variables $(\x,t)$ equations \eqref{eq:nc1}-\eqref{eq:nc2} read
\begin{eqnarray}
  H_t\ +\ \bigl(u - \dot{x}_0(t)\bigr)H_{\x}\ +\ Hu_{\x}\ &=&\ 0, \label{eq:mass} \\
  u_t\ +\ \bigl(u - \dot{x}_0(t)\bigr)u_{\x}\ +\ gH_{\x}\ &=&\ g h'(\x + x_0), \label{eq:moment}
\end{eqnarray}
where $\dot{x}_0(t)$ denotes the derivative of the function $x_0(t)$ with respect to the time $t$. These equations are valid up to the shoreline $\x = 0$. Taking the limit of equation \eqref{eq:mass} as $\x \to 0$ and having in mind the boundary condition \eqref{eq:bc}, one can easily obtain that
\begin{equation}\label{eq:bc2}
  u\bigl(x_0(t),t\bigr)\ =\ \dot{x}_0(t), \qquad \forall t\, \geqslant\, 0.
\end{equation}
On the other hand, the shoreline position $x_0(t)$ was previously found in the form \eqref{eq:taylor}. By substituting the representation \eqref{eq:taylor} into the relation \eqref{eq:bc2}, one can easily compute the shoreline velocity $u\bigl(x_0(t),t\bigr)$ simply by differentiating formally \eqref{eq:taylor}.

\begin{remark}
The computations given above show that only one boundary condition is required in order to determine completely the shoreline motion (in contrast to \cite{Hibberd1979}).
\end{remark}

\begin{remark}
The convergence radius of the series \eqref{eq:taylor} cannot exceed the time $t = t^*$ where the wave breaking occurs. If it occurs in the interior of the fluid domain $x < x_0(t)$, the wave breaking (or \emph{shock formation}, \ie $H_{\x}(\x,t^*) = -\infty$) will be treated numerically by the carefully chosen conservative shock-capturing finite volume scheme. On the other hand, if the wave breaking occurs precisely at the shoreline, it will be treated below in Section~\ref{sec:case3}.
\end{remark}

\begin{remark}
By taking the limit of \eqref{eq:moment} as $\x \to 0$, we can easily compute the wave slope at the shoreline
\begin{equation}\label{eq:slope}
  \left.H_{\x}\right|_{\x = 0}\ =\ h'\bigl(x_0(t)\bigr)\ -\ \frac{1}{g}\left.u_t\right|_{\x = 0}.
\end{equation}
\end{remark}

\subsection{Tangent wave}\label{sec:case2}

\begin{figure}
  \centering
  \includegraphics[width = 0.75\textwidth]{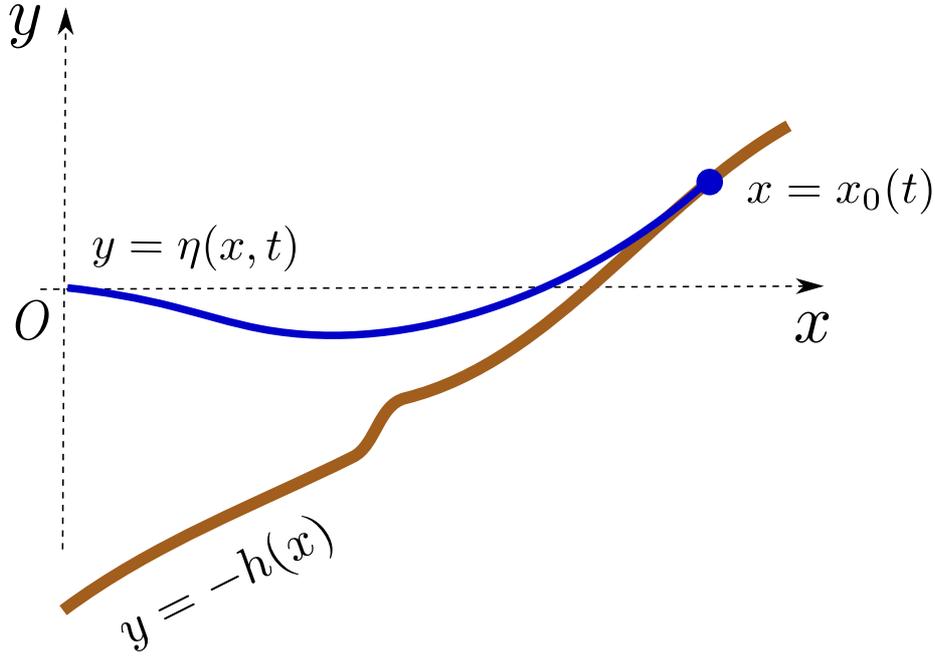}
  \caption{\small\em Sketch of the physical domain when the wave is coming tangent to the shoreline \ie $H_0'\bigl(x_0(t_0)\bigr) = 0$.}
  \label{fig:tangent}
\end{figure}

In this case the initial conditions are chosen such that at the initial moment the free surface geometrically coincides with the tangent drawn at the shoreline. See Figure~\ref{fig:tangent} for the illustration. Analytically this condition reads $H_0'\bigl(x_0(t_0)\bigr) = 0$. It means also that in this particular case the implicit function theorem cannot be applied to determine the shoreline position $x_0(t)$ from the boundary condition \eqref{eq:bc} as we did in the previous Section~\ref{sec:case1}. Consequently, we will adopt another strategy.

Let us consider a more general problem. We assume that initially the free surface is touching the shoreline with the order $p \geqslant 1$, \ie
\begin{equation*}
  H_0^{(i)}\bigl(x_0(t_0)\bigr)\ =\ 0, \quad i = 0,\ldots, p \quad \mbox{ and } \quad
  H_0^{(p + 1)}\bigl(x_0(t_0)\bigr)\ \neq\ 0,
\end{equation*}
where $H_0^{(p)}$ denotes $p$-th derivative. In this case we can seek the function in the form $H(x,t) := \bigl(\H(x,t)\bigr)^p$. Then, in new variables the system \eqref{eq:nc1}-\eqref{eq:nc2} reads
\begin{eqnarray*}
  \H_t\ +\ u\H_x\ +\ \frac{1}{p}\H u_x\ &=&\ 0, \\
  u_t\ +\ uu_x\ +\ pg\H^{p-1}\H_x\ &=&\ gh'(x).
\end{eqnarray*}
Now, similarly to the previous Section, we introduce a new independent variable $\x := x - x_0(t)$ which yields the following system
\begin{eqnarray*}
  \H_t\ +\ \bigl(u - \dot{x}_0(t)\bigr)\H_{\x}\ +\ \frac{1}{p}\H u_{\x}\ &=&\ 0, \\
  u_t\ +\ \bigl(u - \dot{x}_0(t)\bigr)u_{\x}\ +\ pg\H^{p-1}\H_{\x}\ &=&\ gh'(\x + x_0),
\end{eqnarray*}
along with the corresponding transformed initial conditions
\begin{equation*}
  \left.\H\right|_{t = t_0}\ =\ \sqrt[p]{H_0\bigl(\x + x_0(t_0)\bigr)}, \qquad
  \left.u\right|_{t = t_0}\ =\ u_0\bigl(\x + x_0(t_0)\bigr).
\end{equation*}
Finally, by taking the limit as $\x\to 0$ we obtain the following system of Ordinary Differential Equations (ODEs), which governs the shoreline position and velocity as functions of time
\begin{eqnarray}\label{eq:ode1}
  \dot{x}_0 &=& u_0(t), \\
  \dot{u}_0 &=& g h' \bigl(x_0(t)\bigr).\label{eq:ode2}
\end{eqnarray}
The last system has to be completed by appropriate initial conditions. We note that the first equation \eqref{eq:ode1} describes the shoreline kinematics, while the second equation \eqref{eq:ode2} comes from the dynamics.

\begin{remark}
One can notice that the last ODE \eqref{eq:ode2}, which governs the shoreline velocity, can be obtained also from equation \eqref{eq:slope} by remembering that $\left.H_{\x}\right|_{\x = 0} = 0$.
\end{remark}

\begin{remark}
The system of ODEs \eqref{eq:ode1}-\eqref{eq:ode2} is linear and thus, can be solved explicitly if the bottom function $h(x)$ is linear or quadratic in the vicinity of the shoreline $x = x_0(t)$.
\end{remark}

\subsection{Breaking wave}\label{sec:case3}

\begin{figure}
  \centering
  \includegraphics[width = 0.75\textwidth]{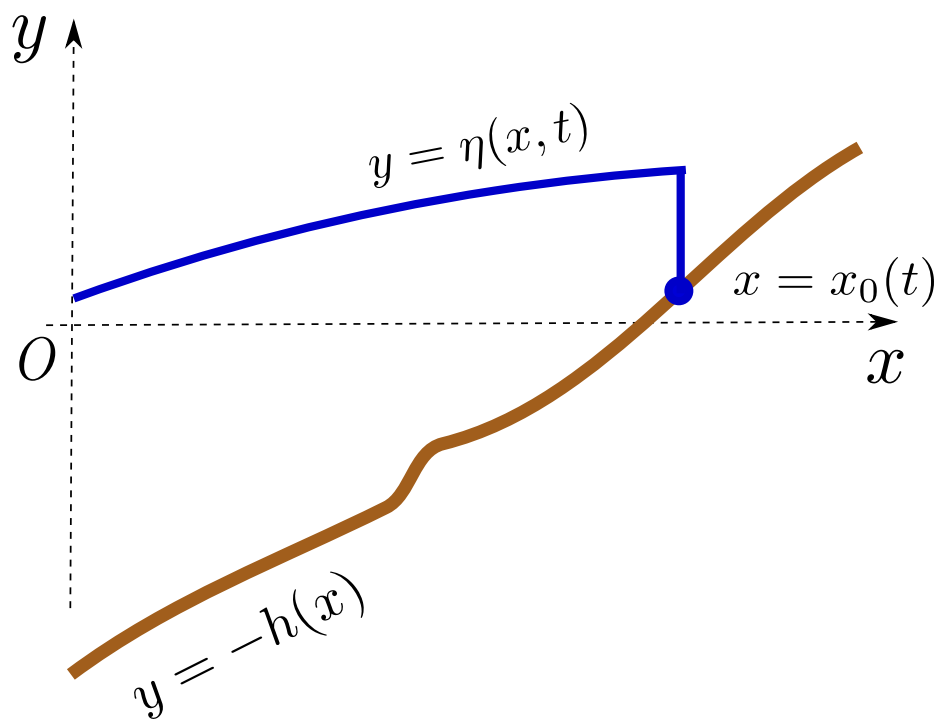}
  \caption{\small\em Sketch of the physical domain when the wave breaking takes place on the shoreline \ie $H_0'\bigl(x_0(t_0)\bigr) = -\infty$.}
  \label{fig:breake}
\end{figure}

\begin{figure}
  \centering
  \includegraphics[width = 0.75\textwidth]{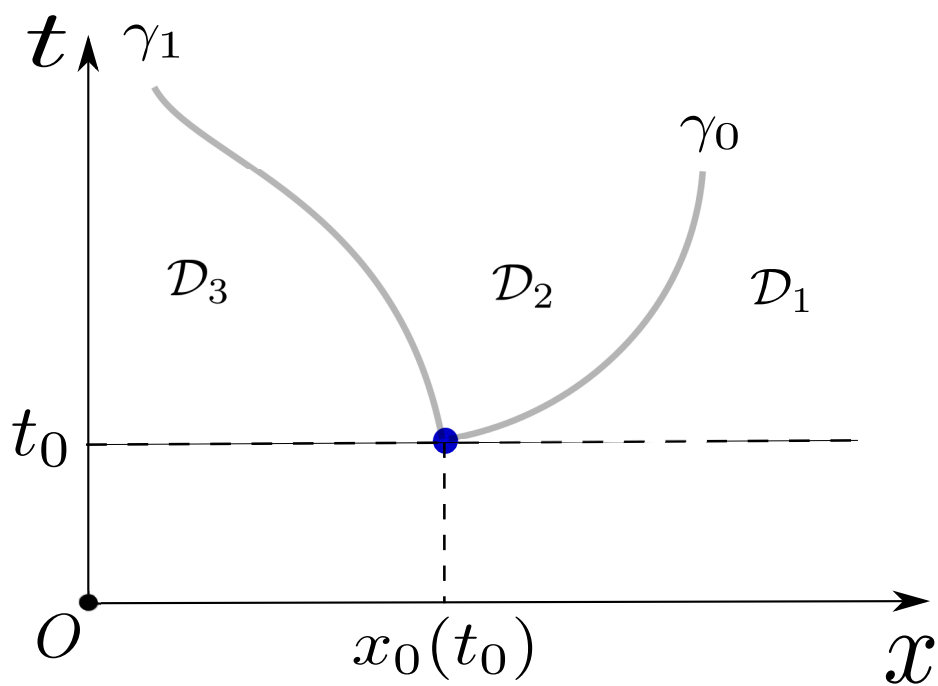}
  \caption{\small\em Solution of the Riemann problem after the wave breaking on the shoreline.}
  \label{fig:breaking2}
\end{figure}

The case when $H_0'\bigl(x_0(t_0)\bigr) = -\infty$ corresponds to the wave breaking which takes place precisely at the shoreline. See Figure~\ref{fig:breake} for the illustration. In this way, we are coming naturally to consider a generalized shoreline Riemann problem. Its classical counterpart was considered earlier in \cite{Stoker1958b, Bellotti2001}. The generalization consists in considering a generic non-constant initial state \eqref{eq:ic}. The evolution of this initial condition in the space-time domain is shown schematically on Figure~\ref{fig:breaking2}. For times $t > t_0$ three distinct domains $\D_{1,2,3}$ can be considered. These domains are separated by two curves $\gamma_{0,1}$ which are defined as:
\begin{itemize}
  \item $\gamma_0 := \bigl(x_0(t), t\bigr)$ is the shoreline trajectory which separates the dry region from the fluid domain. So, in other words we can say that $\left.H(x,t)\right|_{\gamma_0} \equiv 0$ (in case of discontinuous solutions, at least one limit (on the left or on the right) has to vanish). On the other hand, $\left.u(x,t)\right|_{\gamma_0}$ is to be determined.
  \item $\gamma_1 := \bigl(x_1(t), t\bigr)$ is the sonic characteristics $\od{x}{t} = u - c_s$ (a weak discontinuity). The values of the functions $H(x,t)$ and $u(x,t)$ can be determined on the sonic characteristics thanks to the knowledge of initial conditions for $x \leq x_0(t_0)$ and the method of characteristics \cite{Godlewski1990}, which should be used to construct this curve effectively:
  \begin{equation}\label{eq:ic2}
    \left.H(x,t)\right|_{\gamma_1}\ =\ H_1(t), \qquad
    \left.u(x,t)\right|_{\gamma_1}\ =\ u_1(t).
  \end{equation}
\end{itemize}
Once the boundaries $\gamma_{0,1}$ between the sub-domains $\D_i$ are drawn, we can describe their respective content:
\begin{itemize}
  \item $\D_1$ is the dry area, \ie where the total water depth is equal to zero,
  \item $\D_2$ is the transition zone where the solution has to be computed,
  \item $\D_3$ is the unperturbed water state given essentially by the initial conditions.
\end{itemize}
We note however that the precise location of curves $\gamma_{0,1}$ in the $(x,t)$ domain is a part of the shoreline Riemann problem solution. In the sequel we will assume that the unperturbed wave, the sonic curve $\gamma_1$ along with the functions $H_1(t)$, $u_1(t)$ are known.

In order to construct the solution in the sub-domain $\D_2$, we take $t$ and $H$ as independent variables. Hence, $x$ and $u$ become functions of $(t, H)$ at least in some space-time non-empty neighbourhood of the shoreline. The NSWE \eqref{eq:nc1}-\eqref{eq:nc2} in new variables read
\begin{eqnarray}\label{eq:cn1}
  x_t\ &=&\ u\ +\ Hu_H, \\
  x_Hu_t\ -\ Hu_H^2\ +\ g\ &=&\ gx_Hh'(x). \label{eq:cn2}
\end{eqnarray}

\begin{remark}\label{rem:6}
In the new variables the gradients are finite, since on the shoreline we have
\begin{equation*}
  x_H\ \equiv\ \pd{x}{H}\ = \frac{1}{H_x}\ =\ 0, \quad \mbox{ at } \quad t\ =\ t_0.
\end{equation*}
\end{remark}

Thus, the fluid flow in $\D_2$ is described by solutions of the last system of PDEs \eqref{eq:cn1}-\eqref{eq:cn2} with additional conditions posed on boundaries $\gamma_0$ and $\gamma_1$. However, since $\gamma_1$ is a characteristic boundary of multiplicity two, we have to specify one\footnote{Equation \eqref{eq:cn2} degenerates at $t = t_0$ according to Remark~\ref{rem:6}. Thus, only one initial condition has to be specified.} additional condition on it in order to construct a unique analytic solution. In variables $(t, H)$ this condition reads:
\begin{equation}\label{eq:x0h}
  \left.x(t, H)\right|_{t=t_0}\ \equiv\ x_0(H)\ =\ x_0(t_0).
\end{equation}

Now we can construct the formal local analytic solution to the system \eqref{eq:cn1}-\eqref{eq:cn2}. As above, we seek solutions $x(t,H)$ and $u(t,H)$ in the form of power series in $t - t_0$:
\begin{equation}\label{eq:powers}
  x(t,H)\ =\ \sum_{k = 0}^{\infty}\; x_k(H)\; \frac{(t-t_0)^k}{k!}, \qquad
  u(t,H)\ =\ \sum_{k = 0}^{\infty}\; u_k(H)\; \frac{(t-t_0)^k}{k!}.
\end{equation}
This solution is valid only in some neighborhood of $\gamma_1$, where it is expected to be analytic. The lower order coefficient in the expansion of $x(t,H)$ is given by the additional condition \eqref{eq:x0h}. After substituting these expansions into \eqref{eq:cn1}-\eqref{eq:cn2} and setting $t = t_0$ yields
\begin{equation*}
  x_1(H)\ =\ u_0(H)\ +\ Hu_0'(H), \qquad u_0'(H)\ =\ \pm\ \sqrt{\frac{g}{H}}.
\end{equation*}
After integrating the the second equation with respect to $H$ we obtain the following explicit representation
\begin{equation*}
  x_1(H)\ =\ u^*\ \pm\ 3\sqrt{gH}, \qquad u_0(H)\ =\ u^*\ \pm\ 2\sqrt{gH},
\end{equation*}
where $u^*$ appears as an integration constant. In order to determine it we use the initial conditions
\begin{equation*}
  u^*\ =\ u_0\ \pm\ 2\sqrt{gH_0}.
\end{equation*}
For the configuration depicted on Figure~\ref{fig:breake} we choose the value with the positive sign
\begin{equation*}
  u^*\ =\ u_0\ +\ 2\sqrt{gH_0}.
\end{equation*}
The quantity $u^*$ has a precise physical meaning --- it is the instantaneous velocity of the shoreline after the disintegration of the initial discontinuity according to the Riemann problem solution.

After differentiating the system \eqref{eq:cn1}-\eqref{eq:cn2} with respect to time $t$ and setting $t = t_0$ afterwards, we obtain the power series coefficients at the next order
\begin{equation*}
    x_2(H)\ =\ u_1(H)\ +\ Hu_1', \qquad Hu_1'(H)\ -\ \frac34 u_1(H)\ =\ -\frac34 g h'\bigl(x_0(t_0)\bigr).
\end{equation*}
After solving the second differential equation for $u_1(H)$ we obtain the following explicit formulas
\begin{equation*}
    x_2(H)\ =\ \frac{7}{4}u_1^{(1)}H^{\frac34}\ +\ g h'\bigl(x_0(t_0)\bigr), \qquad u_1(H)\ =\ u_1^{(1)}H^{\frac34}\ +\ g h'\bigl(x_0(t_0)\bigr),
\end{equation*}
where $u_1^{(1)}$ is another integration constant. After differentiating the system \eqref{eq:cn1}-\eqref{eq:cn2} $k$ times with respect to $t$ and evaluating the obtained relations at $t = t_0$ we get
\begin{equation*}
  x_{k + 1}\ =\ u_k(H)\ +\ Hu_k'(H), \qquad H u_k'(H)\ -\ \frac34 k u_k(H)\ =\ \sqrt{\frac{H}{4 g}}P_k(H),
\end{equation*}
where $P_k(H)$ is some function determined recursively at every step $k$. The integration of the last differential equation for $u_k'(H)$ yields the following recursion formulas
\begin{eqnarray*}
  x_{k + 1}\ &=&\ \Bigl(1+\frac34 k\Bigr)H^{\frac34 k}u_k^{(1)}\ +\ Q_k(H), \\
  u_k(H)\ &=&\ H^{\frac34 k}\;\biggl[u_k^{(1)}\ +\ \frac{1}{2\sqrt{g}}\int_{t_0}^{t}\frac{P_k(H)}{H^{\frac34 k + \frac12}}\,\ud H\biggr].
\end{eqnarray*}
$Q_k(H)$ is another function of the coefficients obtained on previous steps and $u_k^{(1)}$ are integration constants which can be determined from conditions \eqref{eq:ic2}. Namely, $H_1(t)$ is substituted into the right-hand side of the power series representation of $u(t,H)$ in \eqref{eq:powers}, while $u_1(t)$ is substituted into the left-hand side. Expanding both sides of the equality into a Taylor series in powers of $t - t_0$ and identifying the coefficients in front of equal powers of $t - t_0$ leads the required relations which allow us to determine integration constants $u_k^{(1)}$, $k \geqslant 1$.

From computations made hereinabove we can draw some important conclusions about the properties of the obtained solutions:
\begin{itemize}
  \item By induction we can show that the functions $x(t,H)$ and $u(t,H)$ have the structure
  \begin{equation*}
    x(t,H)\ =\ x^{(0)}(t)\ +\ x_1(t,H),\qquad
    u(t,H)\ =\ u^{(0)}(t)\ +\ u_1(t,H),
  \end{equation*}
  and the functions $x_1(t,H)$ and $u_1(t,H)$ have a known behaviour at the shoreline:
  \begin{equation*}
    \lim_{H\to 0}x_1(t,H)\ =\ 0, \qquad
    \lim_{H\to 0}u_1(t,H)\ =\ 0.
  \end{equation*}
  \item Functions $x^{(0)}(t)$ and $u^{(0)}(t)$ are analytic.
  \item For practical purposes it is easier to compute functions $x^{(0)}(t)$, $u^{(0)}(t)$ by solving the following system of ODEs:
  \begin{equation}\label{eq:ode3}
  \begin{array}{ccccc}
    \dot{x}^{(0)}\ &=&\ u_0,  \qquad x^{(0)}(t_0)\ &=&\ x_0(t_0), \\
    \dot{u}^{(0)}\ &=&\ g h'(x^{(0)}), \qquad u^{(0)}(t_0)\ &=&\ u^*.
  \end{array}
  \end{equation}
  This statement can be checked by expanding the solutions of \eqref{eq:ode3} into a formal series in powers of $t - t_0$ and by identifying the coefficients in front of the equal powers.
  \item As in the previous case, the last system of ODEs is easily and exactly solvable on uniformly sloping or parabolic bottoms when $h'(x) = \const$ or a linear function of its argument.
\end{itemize}

Using the methods described in \cite{Bautin2009}, the following result can be proved
\begin{theorem}[\cite{Bautin2011}]
There exists time $t_2 > t_0$ such that for $\forall t\in[t_0, t_2]$ the series \eqref{eq:powers} converge in the whole sub-domain $\D_2$. Moreover, on the shoreline we have
\begin{equation}\label{eq:cond}
  \left.x_H\right|_{H = 0}\ =\ -\infty, \qquad t_0\ <\ \forall t\ <\ t_2.
\end{equation}
\end{theorem}
By transforming condition \eqref{eq:cond} into the physical space, \ie $\left.H_x\right|_{\gamma_0} = 0$, we can see that the last Theorem has an important
\begin{corollary}
After a wave breaking event taking place exactly on the shoreline, it is the 2\up{nd} scenario (\ie the wave tangent to the shoreline) which is always realized.
\end{corollary}

\begin{remark}
Note the similarity between the systems \eqref{eq:ode1}, \eqref{eq:ode2} and \eqref{eq:ode3}. The difference appears at the level of the initial conditions.
\end{remark}

To summarize the developments made in this section, we constructed a local solution to the generalized shoreline Riemann problem. We underline the local nature of the results presented in this study. All the properties are valid in the vicinity of the shoreline locally in time. However, these solutions turn out to be very useful in numerical computations, where the approximate solution is needed only at the next time step $t_{n + 1} = t_n + \Delta t$, for some appropriately chosen $\Delta t > 0$.


\section{Run-up algorithm}\label{sec:algo}

Now we can briefly describe the run-up algorithm, since all the ingredients have been prepared in the previous sections. Let us consider the 1D grid $\{x_j^{(n)}\}_{j = 1}^{N}$ at time $t = t_n$. The rightmost node $x_N^{(n)}$ corresponds to the moving shoreline at every time layer $t_n$, $n \geqslant 0$. From the definition of the shoreline we know that $H_{N}^{(n + 1)} \equiv 0$. So, we need to determine the current speed $u_N^{(n)}$ and the future position $x_N^{(n + 1)}$ of the shoreline. On every time step we estimate numerically the solution gradient $\delta_x H_{N}^{(n)}$ on the shoreline using a simple finite difference scheme
\begin{equation*}
  \delta_x H_{N}^{(n)}\ :=\ \frac{H_{N}^{(n)} - H_{N-1}^{(n)}}{x_{N}^{(n)} - x_{N-1}^{(n)}}.
\end{equation*}
Then, we choose two numbers $0 < \delta \ll \Delta$. Depending on the value of $\delta_x H_{N}^{(n)}$ the following three scenarii are possible:
\begin{enumerate}
  \item $\delta \leqslant \abs{\delta_x H_{N}^{(n)}} \leqslant \Delta$: we consider that $H'\bigl(x_0(t_n)\bigr) \neq 0$ and we have a regular wave. In order to find the shoreline position $x_0(t)$ at the next time step we use a partial sum (the first $k$ terms) of the series \eqref{eq:taylor}. In practice, the number of terms never exceeds $k = 4$ since it does not lead to the further increase in the accuracy and the numerical results are visually indistinguishable.

  \item $\abs{\delta_x H_{N}^{(n)}} < \delta$: in this case we assume that $H'\bigl(x_0(t_n)\bigr) = 0$ and we have a wave tangent to the bottom at the shoreline. In this case we apply one step of the favourite Runge--Kutta scheme \cite{Bogacki1989} to the system \eqref{eq:ode1}-\eqref{eq:ode2} to obtain the next position of the shoreline.

  \item $\abs{\delta_x H_{N}^{(n)}} > \Delta$: in this case we assume that $H'\bigl(x_0(t_n)\bigr) = -\infty$ and we have a wave breaking event at the shoreline. In this case we also apply one step of a Runge--Kutta scheme to the system \eqref{eq:ode3} to obtain the shoreline position at the next time step $t_{n + 1}$.
\end{enumerate}

\begin{remark}
In the coding practice the numbers $\delta$, $\Delta$ are related to the average spatial discretization step $\langle \Delta x \rangle$ in the following way
\begin{equation*}
  \delta\ :=\ C_1\langle \Delta x \rangle, \qquad
  \Delta\ :=\ \frac{C_2}{\langle \Delta x \rangle},
\end{equation*}
where $C_{1,2}$ are some constants. In the numerical simulations $C_{1,2}$ are chosen such that $\delta = \O(10^{-3})$ and $\Delta = \O(10^{1})$. So, there are four orders of magnitude between the negligibly small and \emph{practically} infinite wave slopes.
\end{remark}

\begin{remark}
In principle, a one step Runge--Kutta scheme could be applied in all three situations. However, in the first case it is not convenient since the righ-hand side contains a term proportional to $H'\bigl(x_0(t)\bigr)$, which has to be evaluated at every stage of the time stepping procedure. It can be done using numerical differentiation, for example, but it can lead eventually to a loss of accuracy and other unnecessary complications. That is why we prefer to exploit instead the analytical solution~\eqref{eq:taylor}.
\end{remark}


\section{Numerical results}\label{sec:res}

In numerical simulations presented below we use a predictor-corrector scheme on an adaptive grid \cite{Shokin2006}. This scheme is monotonicity preserving 2\up{nd} order accurate in space with the reconstruction which is exact for linear solutions \cite{Barth2004}. The CFL condition \cite{Courant1928}, computed using the maximal speed of the characteristics $\lambda_{\pm}$, was set to be equal to $0.95$ in all the computations shown below.

The adaptive grid is constructed using the equidistribution method \cite{Shokin2006}. This algorithm allows to have higher concentrations of grid points which follow the solution extrema \cite{Arvanitis2006} (wave crests, troughs and near the shoreline, depending on its parametrization). The motion of grid points will be shown below for the sake of illustration (see Figure~\ref{fig:grid}).

\subsection{Run-up on a plane beach}

\begin{figure}
  \centering
  \includegraphics[width=0.86\textwidth]{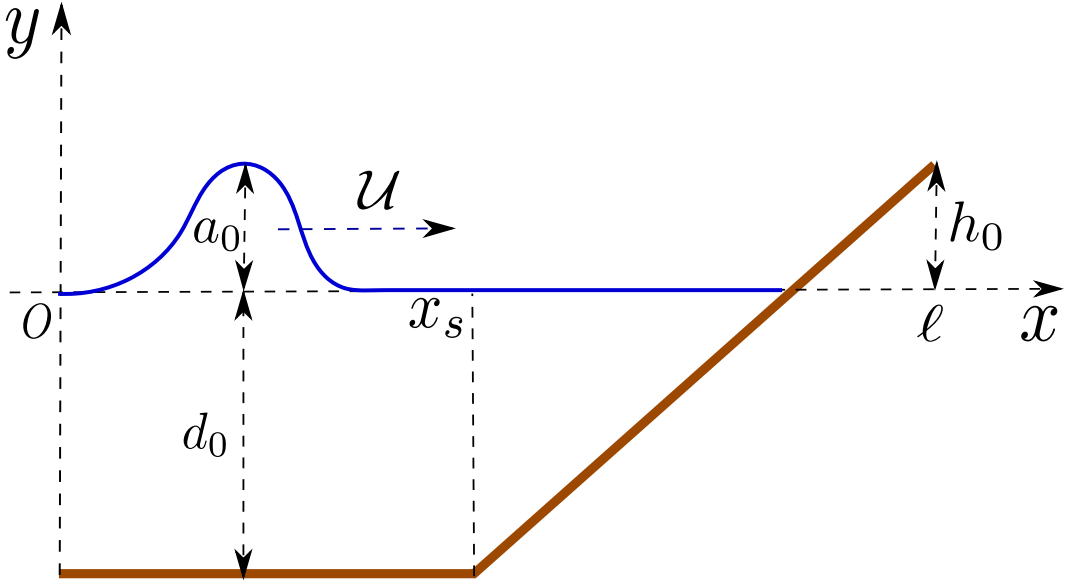}
  \caption{\small\em Sketch of the physical domain corresponding to our numerical simulations of a solitary wave run-up onto a plane beach \cite{Synolakis1986}.}
  \label{fig:beach}
\end{figure}

Let us consider the classical problem of a solitary wave run-up onto a plane beach considered in the PhD thesis of C.~\textsc{Synolakis} \cite{Synolakis1986}. The bottom profile is given by the following function
\begin{equation*}
  y\ =\ -h(x)\ =\ \begin{dcases}
    \;\; -d_0,\; & 0 \leqslant x < x_s, \\
    \;\; -d_0\ +\ (x - x_s)\tan\theta,\; & x_s \leqslant x \leqslant \ell,
  \end{dcases},
  \qquad \ell\ :=\ x_s\ +\ (h_0 + d_0)\cot\theta,
\end{equation*}
where $\theta$ is the bottom slope, $d_0$ is the unperturbed water depth. The sketch of the physical domain is shown on Figure~\ref{fig:beach}. The parameter $h_0$ is the topography height at $x = \ell$, where the wave tank is bounded by a vertical wall. The domain is always chosen such that the wave never touches the boundaries of the computational domain (in order to avoid the influence of boundaries on the numerical results). In the computations presented below, the parameter $h_0$ varied in the range $0.5 \div 1.3$ depending on the wave amplitude and bottom slope.

\begin{figure}
  \centering
  \includegraphics[width = 0.68\textwidth]{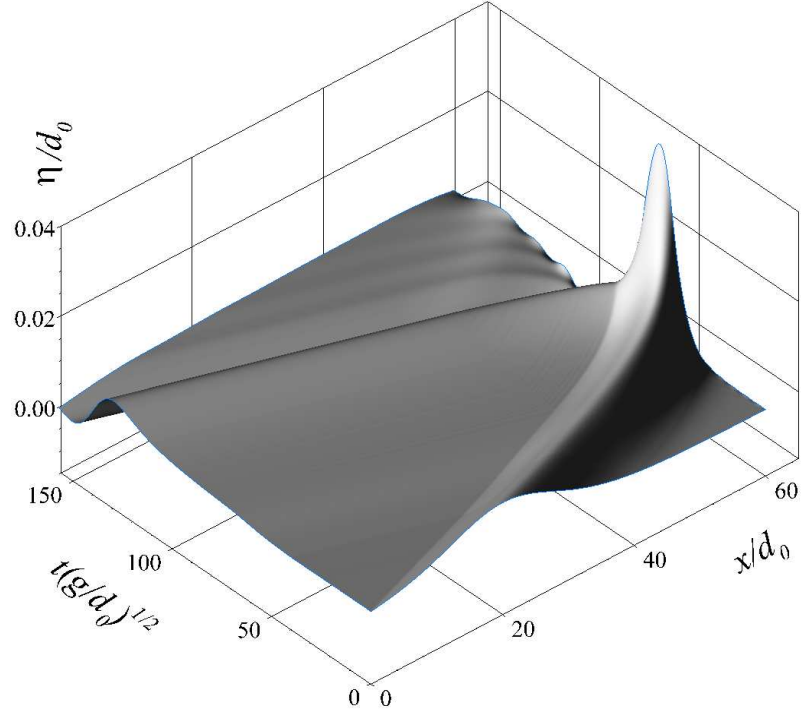}
  \caption{\small\em Space-time plot of the solitary wave run-up with $\theta\ =\ 2.88^\circ$ and $a_0/d_0\ =\ 0.01$.}
  \label{fig:spacetime1}
\end{figure}

\begin{figure}
  \centering
  \includegraphics[width = 0.71\textwidth]{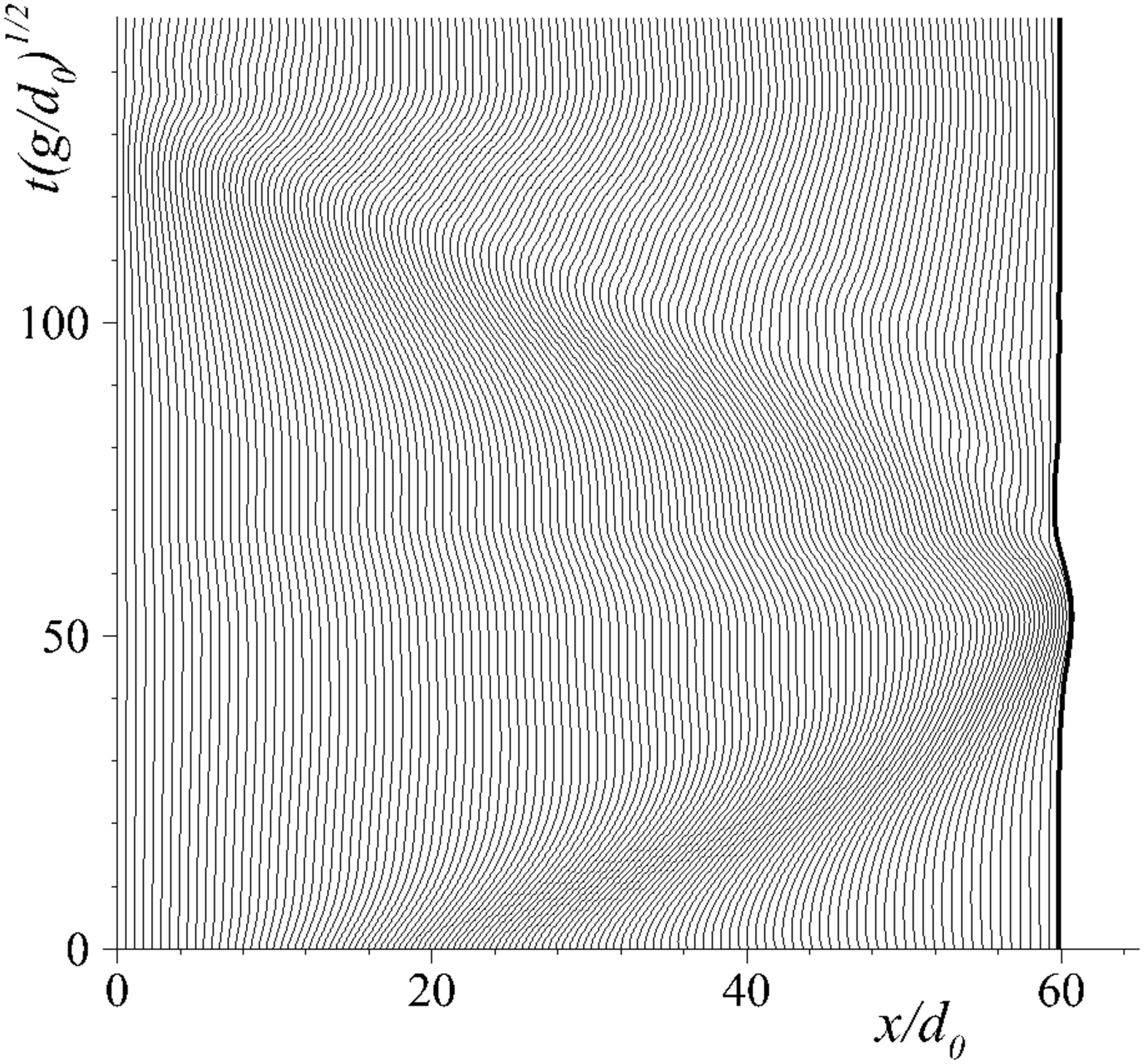}
  \caption{\small\em Trajectories of grid points during a solitary wave run-up with $\theta\ =\ 2.88^\circ$ and $a_0/d_0\ =\ 0.01$. Only every 10\up{th} point is shown for the sake of clarity. This computation corresponds to the space-time evolution shown on Figure~\ref{fig:spacetime1}. The black solid line shows the discrete trajectory of the shoreline point $\{x_N^{(n)}\}$.}
  \label{fig:grid}
\end{figure}

The initial condition was prescribed by the following formulas
\begin{equation}\label{eq:ic0}
  \eta_0(x)\ :=\ a_0\,\sech^2\Bigl(\frac{\sqrt{3 a_0 g}}{2\,d_0\U}(x - x^{0})\Bigr), \quad H_0(x)\ :=\ h(x)\ +\ \eta_0(x), \quad u_0(x)\ :=\ \U\frac{\eta_0(x)}{H_0(x)},
\end{equation}
where $a_0$ is the solitary wave amplitude, $x^{0}$ is the initial position of the wave crest and $\U := \sqrt{g(d_0 + a_0)}$ is the wave celerity in the fully nonlinear, weakly dispersive formulation \cite{Serre1953a, Green1976, Dutykh2011a}. In the numerical results shown below we chose $x^{0}/d_0 := 20$ and $x_s/d_0 := 40$. A sample simulation of a wave run-up for $a_0/d_0\ =\ 0.01$ and the bottom slope $\theta\ =\ 2.88^\circ$ is shown on Figure~\ref{fig:spacetime1}. The motion of every 10\up{th} grid point is represented on Figure~\ref{fig:grid} in order to show the work of the adaptive strategy for the grid motion.

Synolakis proposed an empirical law to determine the maximal run-up $\Ru$ of a solitary wave on a plane beach \cite{Synolakis1987}:
\begin{equation}\label{eq:rup}
  \frac{\Ru}{d_0}\ =\ \underbrace{2.831\sqrt{\cot\theta}\;\Bigl(\frac{a_0}{d_0}\Bigr)^{\frac54}}_{S}.
\end{equation}
This condition was obtained for non-breaking waves under the additional assumption
\begin{equation}\label{eq:sy}
  \frac{a_0}{d_0}\ >\ \bigl(0.288\tan\theta\bigr)^2.
\end{equation}
The breking criteria during the wave run-up was given in \cite{Synolakis1987}:
\begin{equation*}
  \frac{a_0}{d_0}\ <\ 0.82\;(\tan\theta)^{\frac{10}{9}},
\end{equation*}
and during the run-down in \cite{Pedersen1983}:
\begin{equation}\label{eq:ped}
  \frac{a_0}{d_0}\ <\ 0.479\;(\tan\theta)^{\frac{10}{9}}.
\end{equation}
Please, note that the wave breaking is more likely to happen during the run-down than during the run-up (since $0.479 < 0.82$).

\begin{figure}
  \centering
  \includegraphics[width=0.69\textwidth]{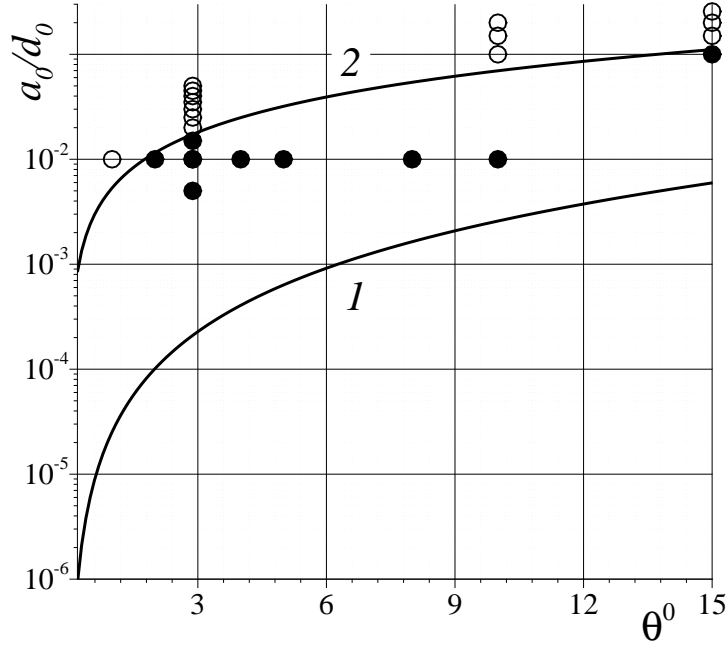}
  \caption{\small\em Applicability range of formula \eqref{eq:rup} bounded from below by condition \eqref{eq:sy} and from above by the wave breaking criterium \eqref{eq:ped}. Numerical experiments which lie in the range of applicability of \eqref{eq:rup} are shown with filled circles ($\bullet$). All the others, which are outside, are shown with empty circles ($\circ$).}
  \label{fig:breaking}
\end{figure}

\begin{figure}
  \centering
  \includegraphics[width=0.78\textwidth]{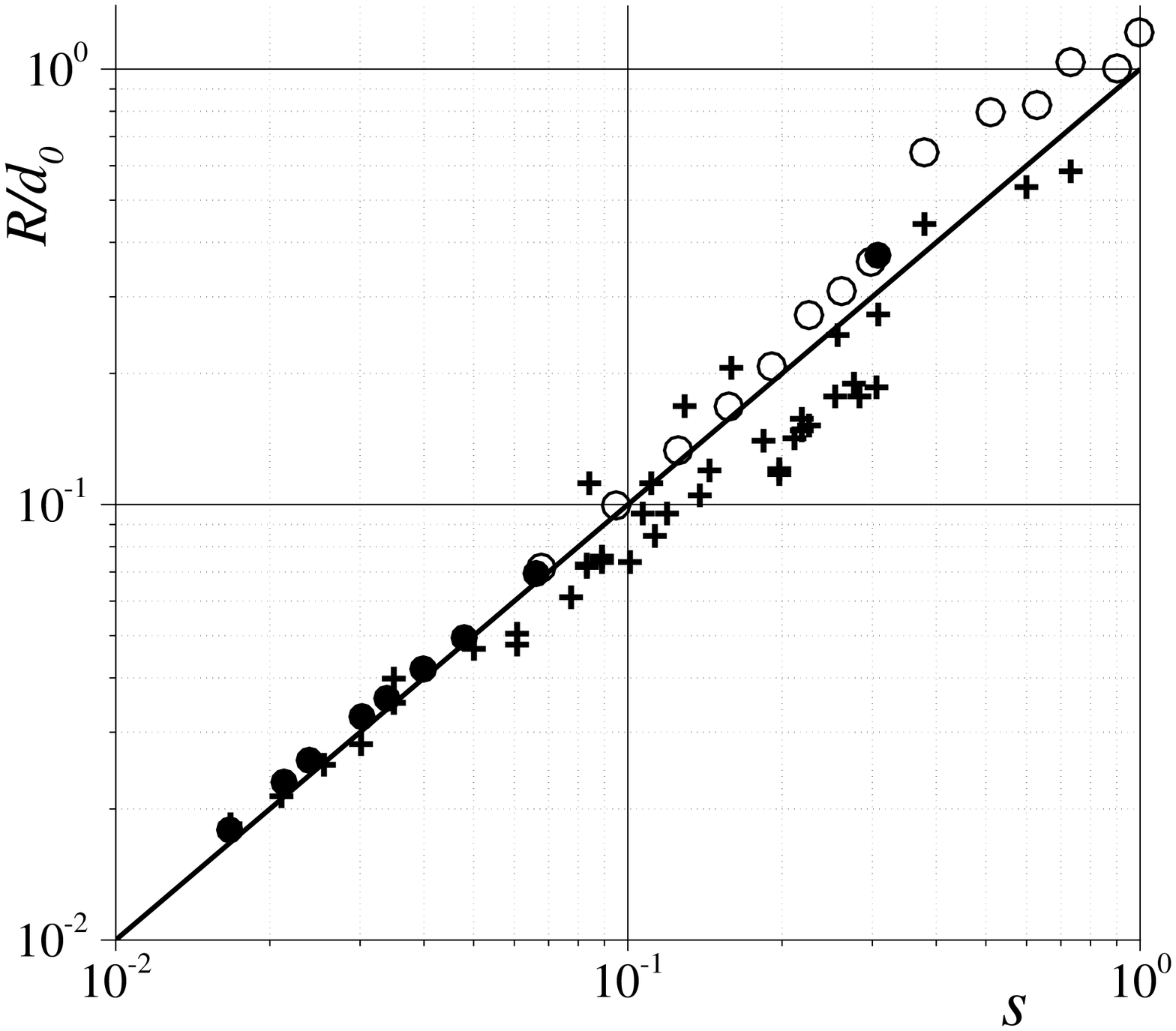}
  \caption{\small\em Comparison of maximal run-up values with the experimental data ($\boldsymbol{+}$) \cite{Synolakis1987} and the analytical prediction \eqref{eq:rup} (straight solid line). The horizontal variable $S$ denotes the right hand side of the equation \eqref{eq:rup}.}
  \label{fig:comp}
\end{figure}

We performed a series of numerical experiments (23 in total) for various values of the wave amplitude $a_0/d_0 = 0.005 \div 0.256$ and bottom inclinations $\theta = 1^\circ\div 15^\circ$. All these experiments are shown with filled ($\bullet$) or empty ($\circ$) circles on Figure~\ref{fig:breaking} depending on whether they fall ($\bullet$) into the applicability range of formula \eqref{eq:rup} or not ($\circ$). Comparisons with the analytical prediction \eqref{eq:rup} and experimental data \cite{Synolakis1987} show an excellent agreement with our numerical results presented on Figure~\ref{fig:comp}. Moreover, these results seem to indicate that the applicability range of formula \eqref{eq:rup} has been seriously underestimated.

\subsection{Run-up on a curvilinear beach}

\begin{figure}
  \centering
  \includegraphics[width=0.69\textwidth]{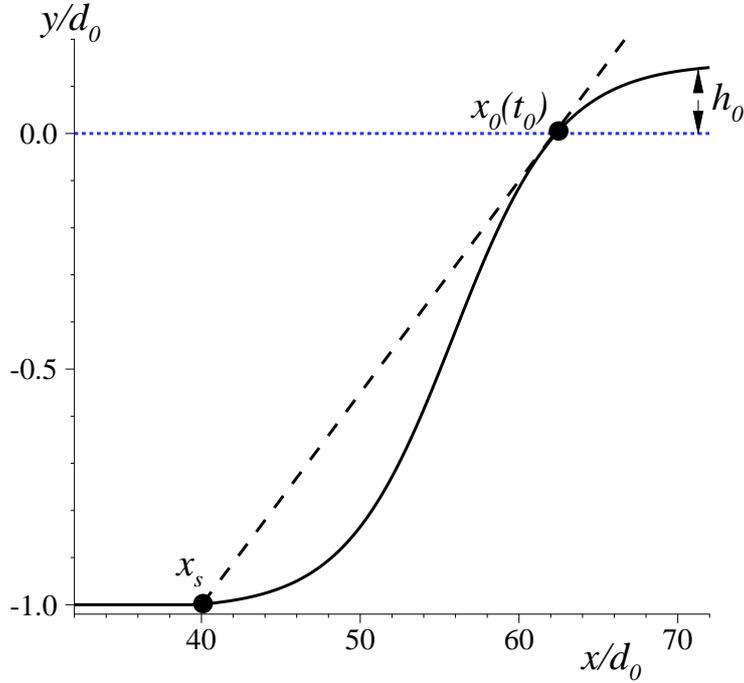}
  \caption{\small\em Bottom profile given by formula \eqref{eq:bot} (black solid line) for $\theta = 5^\circ$. Dashed line shows the corresponding linearized bottom profile \eqref{eq:linear}.}
  \label{fig:bottoms}
\end{figure}

It was explained above that our run-up algorithm becomes particularly simple and elegant for linear or parabolic bottoms (in the vicinity of the shoreline) considered so far. In order to show the performance of the algorithm in general cases, we consider the following general bathymetry:
\begin{equation}\label{eq:bot}
  y\ =\ -h(x)\ =\ \begin{dcases}
  \;\; -d_0\; & 0 \leqslant x \leqslant x_s, \\
  \;\; \frac{h_0 - d_0}{2}\ +\ \Bigl(h_{\infty}\ -\ \frac{h_0 - d_0}{2}\Bigr)\tanh\bigl(\kappa(x - \xi)\bigr), \; & x_s < x \leqslant \ell, \\
  \end{dcases},
\end{equation}
where $d_0$ is the unperturbed water depth in the leftmost point, $h_0$, $h_{\infty}$ are the topography heights at $x = \ell$ and $x \to \infty$ correspondingly. Parameters $\kappa$, $\xi$ and $\ell$ are defined as
\begin{equation*}
  \kappa\ :=\ \frac{\tan\theta}{h_{\infty}\ -\ \frac{h_0 - d_0}{2}}, \qquad
  \xi\ :=\ x_s\ +\ \frac{1}{2\kappa}\ln\frac{h_{\infty} + d_0}{h_{\infty} - h_0}, \qquad
  \ell\ :=\ 2\xi\ -\ x_s,
\end{equation*}
where $\theta$ is the maximal bottom slope which is reached at $x = \xi$. In the computations presented in this Section we use the values $d_0 = 1.0$, $h_{\infty} = 0.15$ and $h_0 = 0.14$. Moreover, we take $x_s/d_0 = 40.0$ and the solitary wave is placed initially at $x^0/d_0 = 20.0$. The unperturbed position of the shoreline $x_0(t_0)$ for the bottom slope \eqref{eq:bot} is
\begin{equation*}
  x_0(t_0)\ =\ \xi\ +\ \frac{1}{2\kappa}\ln\Bigl(1\ +\ \frac{1 - h_0}{h_{\infty}}\Bigr), \qquad \xi\ <\ x_0(t_0)\ <\ \ell.
\end{equation*}
For the sake of comparison we considered also an \emph{equivalent} linearized bottom profile given naturally by the secant joining the point $(x_s, -d_0)$ with $(x_0(t_0), 0)$:
\begin{equation}\label{eq:linear}
  y\ =\ -h(x)\ =\ \begin{dcases}
    \;\; -d_0\; & 0 \leqslant x \leqslant x_s, \\
    \;\; \frac{x - x_0(t_0)}{x_0(t_0) - x_s}, \; & x_s < x \leqslant \ell
  \end{dcases}.
\end{equation}
Two profiles of the curvilinear bathymetry given by formula \eqref{eq:bot} along with corresponding linearized profiles \eqref{eq:linear} are depicted on Figure~\ref{fig:bottoms}. The numerical values of several bottom characteristics for various choices of the parameter $\theta$ are given in Table~\ref{tab:bottom}.

\begin{table}
  \centering
  \begin{tabular}{||c||c|c||c|c|c||c||}
  \hline\hline
   $\theta$ & $\xi$ & $x_0(t_0)$ & $\ell$ & $\tilde{\theta}_{0}$ & $\theta_{s}$ & $\tilde{\theta}$ \\ 
   \hline\hline
   3 & 66.25 & 76.81 & 92.51 & 1.35 & 0.10 & 1.56 \\
   5 & 55.73 & 62.05 & 71.46 & 2.26 & 0.17 & 2.60 \\
   7 & 51.20 & 55.71 & 62.41 & 3.16 & 0.24 & 3.64 \\
   9 & 48.69 & 52.18 & 57.38 & 4.08 & 0.31 & 4.69 \\
  11 & 47.08 & 49.93 & 54.16 & 5.00 & 0.38 & 5.75 \\
  \hline\hline
  \end{tabular}
  \bigskip
  \caption{\small\em Dependence of the linearized bottom characteristics on the parameter $\theta$ (curvilinear bottom slope at $x = \xi$) from formula \eqref{eq:bot}. The parameters $\tilde{\theta}_{0}$ is the curvilinear bottom slope at $x = x_0(t_0)$; $\theta_{s}$ is the curvilinear bottom slope at $x = x_s$. The slope of the linearized bottom profile is $\tilde{\theta}$.}
  \label{tab:bottom}
\end{table}

\begin{figure}
  \centering
  \subfigure[Curvilinear bottom]{
  \includegraphics[width=0.70\textwidth]{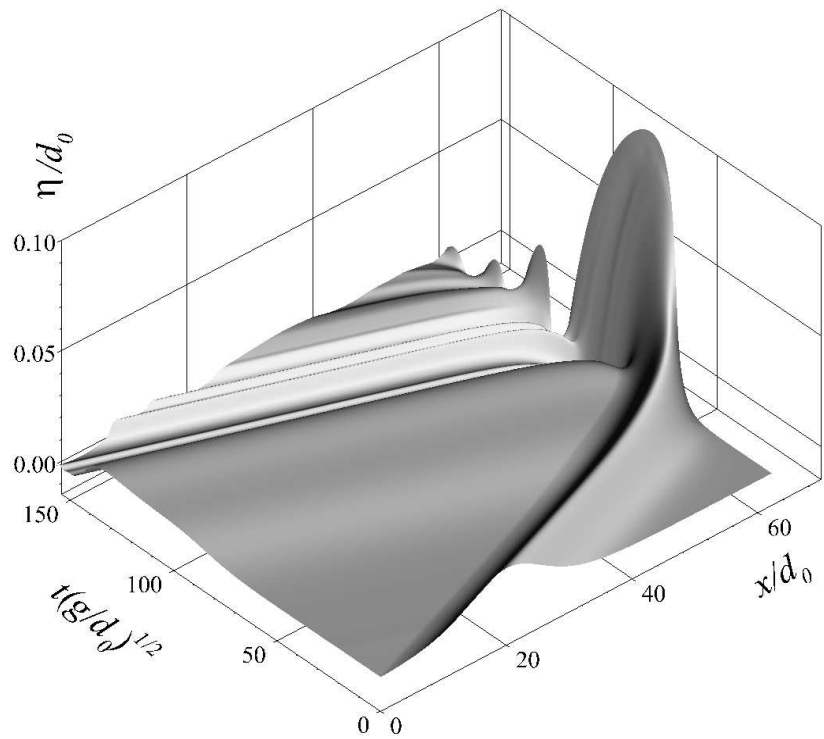}}
  \subfigure[Linearized bottom]{
  \includegraphics[width=0.70\textwidth]{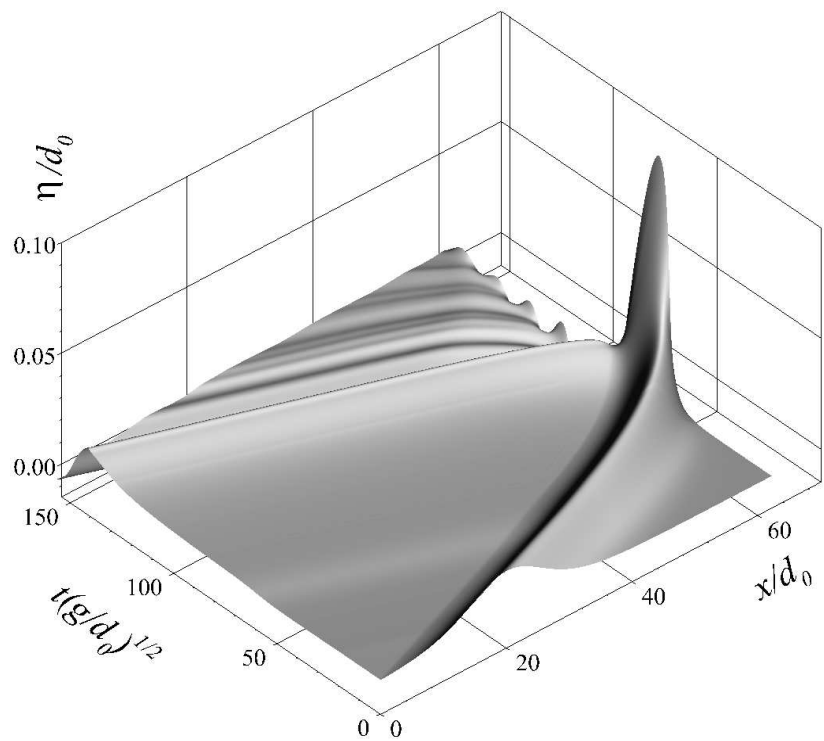}}
  \caption{\small\em Space-time plots of the free surface evolution for $a_0/d_0 = 0.02$ and $\theta = 5^\circ$ for the curvilinear (a) and corresponding linearized (b) bottom profiles.}
  \label{fig:quad}
\end{figure}

We performed in total $20 = 5\times 2\times 2$ numerical experiments of a solitary wave run-up \eqref{eq:ic0} for five values of the bottom slope $\theta$, two values of the incident wave amplitude and two shapes of the bottom. All the values of these parameters along with computed maximal run-ups $\Ru/a_0$ are reported in Table~\ref{tab:rup}. It turns out that maximal run-up in all cases considered in this study are slightly higher for the curvilinear bottom \eqref{eq:bot}. This result shows one more time that the precise knowledge of the local bathymetric features is of capital importance for the accurate prediction of the wave run-up. For the sake of comparison we show on Figure~\ref{fig:quad} the space-time plots of the free surface elevation ($a_0/d_0 = 0.02$ and $\theta = 5^\circ$) for both bottom profiles. In particular, the curvilinear bottom induces on Figure~\ref{fig:quad}(\textit{a}) a higher and much longer first wave along with more pronounced secondary oscillations.

\begin{table}
 \centering
  \begin{tabular}{||c|c|c||c|c||}
  \hline\hline
   & \multicolumn{4}{c||}{$\Ru/a_0$} \\
  \cline{2-5}
  $\theta_0$&  \multicolumn{2}{c||}{\textit{Curvilinear bottom} \eqref{eq:bot}}&  \multicolumn{2}{c||}{\textit{Linearized bottom} \eqref{eq:linear}} \\ \cline{2-5}
  & $a_0/d_0\ =\ 0.01$& $a_0/d_0\ =\ 0.02$ & $a_0/d_0\ =\ 0.01$ & $a_0/d_0\ =\ 0.02$ \\ \hline
  $3^\circ$  & 5.62 & 6.34 & 5.58 & 6.30 \\
  $5^\circ$  & 4.40 & 5.15 & 4.37 & 5.11 \\
  $7^\circ$  & 3.71 & 4.36 & 3.74 & 4.42 \\
  $9^\circ$  & 3.26 & 3.90 & 3.35 & 3.94 \\
  $11^\circ$ & 2.95 & 3.52 & 3.07 & 3.60 \\
  \hline\hline
  \end{tabular}
  \bigskip
  \caption{\small\em Dependence of the maximal wave run-up on the bottom slope, incident wave amplitude and the bottom shape.}
  \label{tab:rup}
\end{table}

\section{Discussion}\label{sec:concl}

The main conclusions and perspectives of this study are outlined below.

\subsection{Conclusions}

In this article a novel asymptotic solution to the \acs{NSWE} is derived. This solution is directly inspired by analogous problems arising in the Gas Dynamics (\eg gas outflow into the vacuum, \cite{Bautin2005}). So, some methods of compressible fluid mechanics have been transposed to shallow water flows thanks to the mathematical analogy between the governing equations. We would like to stress out that our asymptotic solution is valid for general bathymetries in contrast to several other analytical investigations limited exclusively to the sloping beach case \cite{CG58, CWY}. A new run-up algorithm was proposed based on this deeper analytical knowledge of \acs{NSWE} solutions structure in the vicinity of the wet/dry interface transition. Moreover, this algorithm uses moving grids in order to mesh the fluid domain only which leads to significant savings in terms of the CPU time. The usage of this algorithm was illustrated on several realistic examples. Finally, we note that the proposed run-up algorithm is particularly simple and elegant for uniformly sloping or parabolic bottoms.

We believe that the methodology presented in this study will be especially important for higher-order Finite Volumes (FV) or Discontinuous Galerkin (DG) discretizations \cite{Bookhove2005} where the high accuracy is required throughout the whole domain. The tools developed here allow to make a mixed numerical/analytical zoom on the shoreline kinematics and dynamics.

\subsection{Perspectives}

The first natural extension consists in generalizing the present algorithm to 3D (2DH) flows on structured \cite{MarcheBonneton2007, Medeiros2013} or unstructured \cite{Dutykh2009a, Delis2011} meshes. Another important question to be investigated in future studies is the interaction of the proposed run-up algorithm with terms not considered in the present formulation. For instance, one could think about the inclusion of the friction effects at the bottom \cite{Vazquez-Cendon1999, Bohorquez2010, Antuono2012} or, even more importantly, of the non-hydrostatic effects \cite{Dutykh2011e, Dutykh2011d, Tissier2011, Shi2012}. A special attention might be needed since Boussinesq-type equations are known to be prone to develop numerical instabilities in general \cite{Lovholt2009} and particularly in the vicinity of the shoreline \cite{Bellotti2002}.


\subsection*{Acknowledgments}
\addcontentsline{toc}{subsection}{Acknowledgments}

This research was supported by RSCF project No 14-17-00219. D.~Dutykh acknowledges the support of the Dynasty Foundation as well as the hospitality of the Institute of Computational Technologies SB RAS during his visit in October 2014. This work was finalized later at the Institute f\"ur Analysis, Johannes Kepler Universit\"at Linz whose hospitality is equally acknowledged. D.~\textsc{Mitsotakis} was supported by the Marsden Fund administered by the Royal Society of New Zealand. All the authors would like to thank the anonymous Referee for very helpful suggestions on our manuscript.


\addcontentsline{toc}{section}{References}
\bibliographystyle{abbrv}
\bibliography{biblio}

\end{document}